\documentclass[11pt,a4paper]{article}
\usepackage[pdftex]{graphicx}     
\usepackage{enumerate}
\usepackage{natbib}
\usepackage{url} % not crucial - just used below for the URL 
\usepackage{amsmath,amsfonts,enumerate,color,verbatim,booktabs,amssymb}
\usepackage{algorithm}
\usepackage{microtype}
\usepackage{xspace}
\usepackage{color}
\usepackage{mathtools}
\usepackage[normalem]{ulem} 

\usepackage[english]{babel}
\graphicspath{{graphics/}}

\renewcommand{\vec}[1]{\boldsymbol{#1}}
\newcommand{\yvec}{\vec{y}}
\newcommand{\thetavec}{\vec{\theta}}
\newcommand{\dvec}{\vec{d}}

\newcommand{\muvec}{\vec{\mu}}

\newcommand{\mvec}{\vec{m}}
\newcommand{\vvec}{\vec{v}}
\newcommand{\wvec}{\vec{w}}
\newcommand{\lvec}{\vec{\ell}}
\newcommand{\Var}{\mbox{Var}}
\newcommand{\Cov}{\mbox{Cov}}
\newcommand{\eps}{\varepsilon}
\newcommand{\K}{K}
\newcommand{\E}{\text{E}}
\newcommand{\D}{\mathcal{D}}
\renewcommand{\L}{\mathcal{L}}
\newcommand{\LN}{\text{LN}}
\newcommand{\N}{\text{N}}
\newcommand{\Cat}{\text{Cat}}
\newcommand{\Ga}{\text{Ga}}
\newcommand{\Beta}{\text{Beta}}
\newcommand{\Muller}{\text{M{\"u}ller}\xspace}
\newcommand{\indep}{\overset{indep}{\sim }}

\usepackage{booktabs}
\graphicspath{{graphics/}}

\title{Efficient construction of Bayes optimal designs for stochastic process models}
\author{Colin S. Gillespie\thanks{email:
    \texttt{colin.gillespie@newcastle.ac.uk}} \and\ Richard J. Boys}
\date{School of Mathematics, Statistics and Physics, Newcastle University,\\
  Newcastle upon Tyne, NE1 7RU, UK}

\begin{document}
\maketitle
\begin{abstract}
  Stochastic process models are now commonly used to analyse complex
  biological, ecological and industrial systems. Increasingly there is
  a need to deliver accurate estimates of model parameters and assess
  model fit by optimizing the timing of measurement of these
  processes. Standard methods to construct Bayes optimal designs, such
  the well known \Muller algorithm, are computationally intensive even
  for relatively simple models. A key issue is that, in determining
  the merit of a design, the utility function typically requires
  summaries of many parameter posterior distributions, each determined
  via a computer-intensive scheme such as MCMC. This paper describes a
  fast and computationally efficient scheme to determine optimal
  designs for stochastic process models. The algorithm compares
  favourably with other methods for determining optimal designs and
  can require up to an order of magnitude fewer utility function
  evaluations for the same accuracy in the optimal design solution. It
  benefits from being embarrassingly parallel and is ideal for running
  on multi-core computers. The method is illustrated by determining
  different sized optimal designs for three problems of increasing
  complexity.
\end{abstract}


\section{Introduction}
\label{sec:intro}
Stochastic process models are increasingly used to describe the
dynamic evolution of a complex system containing different interacting
species.  Applications appear in many areas such as biology, ecology,
pharmacokinetics and industry; see, for example, \cite{Henderson09},
\cite{CookGG2008}, \cite{RyanDP2015} and \cite{KhatabAEDD17}. Designed
experiments can be very useful to the practitioner as they allow them
to learn about models and their parameters in an efficient way. For
example, in systems biology, an experimenter might build a stochastic
kinetic model for their biological system; these models are typically
described through a series of reactions between the species, with each
reaction depending on an unknown stochastic rate constant. Data are
then collected with the aim of estimating these constants and
assessing model fit.  Clearly scheduling the timing of say~$k$
observations within a $(0,T)$ experimental time period has the
potential to yield much more accurate inferences than say just
observing the process at $k$~times on a regular grid.

We consider designs in which the stochastic process is observed on $k$
occasions, at times $\dvec=(t_1,\ldots,t_k)$. In general, the merit of
a particular design~$\dvec$ is captured through a utility function
$u(\dvec,\yvec,\thetavec)$, where $\yvec$ are data that might be
observed at times $\dvec$ when the model parameter is~$\thetavec$. In
this paper we focus on utility functions based on the posterior
distribution of~$\thetavec$, namely, the posterior generalised
precision
\begin{equation}
u(\dvec,\yvec)=1/\text{det}\{\Var(\thetavec|\yvec,\dvec)\}
\label{eq:genprec}
\end{equation}
and its logarithm. These utilities have an intuitive motivation and
are appropriate if the posterior is unimodal and without substantial
skewness or kurtosis. Note that these utility functions do not depend
on the model parameter~$\thetavec$. Other popular utility functions
(not used here) that do depend on the model parameter~$\thetavec$ are
those based on the self-information loss, absolute error loss and
squared error loss \citep{OverstallMD18}.

As the choice of design must be made before observing data, designs
should be assessed by their expected utility
\[
u(\dvec)=E_{\yvec}\{u(\dvec,\yvec)\}=\int_{\yvec}u(\dvec,\yvec)\pi(\yvec|\dvec)\,d\yvec,
\]
where
$\pi(\yvec|\dvec)=\int_{\thetavec}\pi(\yvec|\dvec,\thetavec)\pi(\thetavec)d\thetavec$
is the prior predictive density of the unobserved data,
$\pi(\yvec|\dvec,\thetavec)$ is the density of the unobserved
data~$\yvec$ when using design~$\dvec$ and the model parameter
is~$\thetavec$, and $\pi(\thetavec)$ is the prior density describing
uncertainty in the model parameter. Therefore the optimal design
$\dvec^*$ over a design space~$\D$ is given by
\[
\dvec^*=\arg\max_{\dvec\in\D} u(\dvec).
\]
Unfortunately the expected utility $u(\dvec)$ is rarely analytically
tractable and so computational schemes are needed. As standard Monte
Carlo integration methods are also not feasible for non-trivial
problems, \cite{Muller1999} proposed a Monte Carlo Markov chain (MCMC)
approach. Although the general \Muller scheme allows for the utility
function to depend on~$\thetavec$, here we focus on utility functions
which depend only on the design~$\dvec$ and potential
observations~$\yvec$. In this scenario the \Muller scheme targets the
density
\[
h(\dvec,\yvec)\propto u(\dvec,\yvec)\pi(\yvec|\dvec)
\] 
using the MCMC scheme in Algorithm~\ref{alg:Muller} (with $J=1$).  The
key feature of this scheme is that the marginal distribution over
$\yvec$ is proportional to the expected utility $u(\dvec)$. Therefore
the optimal design can be estimated as the mode of the empirical
marginal for~$\dvec$, obtained from the MCMC sample. Note that this
scheme mixes over design space by proposing moves using a proposal
distribution~$q$ for designs. Also realisations from the prior
predictive distribution of the unobserved data $\pi(\yvec|\dvec)$ are
obtained straightforwardly by first simulating a parameter
value~$\thetavec$ from the prior distribution and then data~$\yvec$
from the stochastic model.
\begin{algorithm}[t]
 \begin{tabular}{@{}ll@{}}
   {\small 1:} & Initialise $\dvec^0$, simulate $\yvec^0_1,\ldots,\yvec^0_J\indep\pi(\yvec|\dvec^0)$\\
   {\small 2:} & Calculate $u^0=\prod_{j=1}^Ju(\dvec^0,\yvec^0_j)$ \\
   {\small 3:} & \textbf{for} $i=1$ to $N$ \textbf{do}\\
   {\small 4:} & \quad Propose $\dvec^*\sim q(\dvec|\dvec^{i-1})$\\
               & \qquad and simulate $\yvec^*_1,\ldots,\yvec^*_J\indep\pi(\yvec|\dvec^*)$\\
   {\small 5:} & \quad Calculate $u^*=\prod_{j=1}^Ju(\dvec^*,\yvec^*_j)$ \\
   {\small 6:} & \quad \textbf{if} $U(0,1)<\min\left(1,u^*q(\dvec^{i-1}|\dvec^*)/\{u^{i-1}q(\dvec^*|\dvec^{i-1})\}\right)$ \\
   {\small 7:} & \qquad \textbf{then} $u^i=u^*$, $\dvec^i=\dvec^*$ \textbf{else} $u^i=u^{i-1}$, $\dvec^i=\dvec^{i-1}$\\
   {\small 8:} & \textbf{end} \\
  \caption{MCMC algorithm by \citet{Muller1999}}\label{alg:Muller}
  \end{tabular}
\end{algorithm}
\Muller suggested that this mode might be identified more easily by
using $J>1$ replicates from the prior predictive distribution. Here
the MCMC scheme target becomes
\[
h_J(\dvec,\yvec_1,\ldots,\yvec_J)\propto\prod_{j=1}^J u(\dvec,\yvec_j)\pi(\yvec_j|\dvec).
\]
The marginal for $\dvec$ is now proportional to $u(\dvec)^J$ and so,
as $J$ increases, the variance of this marginal is reduced and the
mode is identified more easily. However, a significant problem with
this algorithm is that, for large~$J$, the computational burden
becomes prohibitive and the algorithm risks getting stuck in a local
mode.

\cite{Amzal2006} have developed a particle-based approach to target
$h_J(\dvec)$. Their method begins by first constructing a list of
increasing values of $J$: $J_1<J_2<\cdots<J_M$. The main loop in their
Resampling-Markov algorithm starts with a sample drawn approximately
from $h_{J_{m-1}}(\dvec)$ that is then resampled and enriched by a
Markov step (with proposal distribution $q_{MH}$) to become an
approximated sample from $h_{J_m}(\dvec)$. The algorithm for a
particular choice of~$J$ is given in Algorithm~\ref{alg:Amzal}. It has
many strengths over the standard \Muller algorithm, particularly in
the way it adaptively searches for the optimal design.

More recently, algorithms have been developed which are aimed at
determining high-dimensional designs: the approximate coordinate
exchange (ACE) algorithm \citep{OverstallW17} and the induced natural
selection heuristic (INSH) algorithm \citep{PriceBRT18}. The ACE
algorithm can be implemented using the R package \texttt{acebayes}
\citep{Overstall2017}. R~code for the INSH algorithm has been provided
to us by the authors. All code for the examples in this paper can be
found at \verb#https://github.com/csgillespie/expt_design#.

\begin{algorithm}[t]
 \begin{tabular}{@{}ll@{}}
   {\small 1:} & \textbf{for} $i=1$ to $N$ \textbf{do}\\
   {\small 2:} & \quad Simulate $\yvec_{ij}^{m-1}\indep\pi(\yvec|\dvec_i^{m-1})$,\\
               & \qquad $j=J_{m-1}+1,\ldots,J_m$\\
   {\small 3:} & \quad Calculate $w_i^m= w_i^{m-1} \prod_{j=J_{m-1}+1}^{J_m} u(\dvec_i^{m-1},\yvec^{m-1}_{ij})$ \\
   {\small 4:} & \textbf{end} \\
   {\small 5:} & Simulate $(\ell_1,\ldots,\ell_N)\sim M(N,\wvec^m)$ \\
   {\small 6:} & \textbf{for} $i=1$ to $N$\quad Set $\widehat{\dvec}_i=\dvec_{\ell_i}$ and 
                 $\widehat{u}_i^m=\widehat{u}_{\ell_i}^{m-1}\times w_{\ell_i}^m$ \\
   {\small 7:} & \textbf{for} $i=1$ to $N$ \textbf{do}\\
   {\small 8:} & \quad Simulate $\overline{\dvec}_i^m\sim q_{MH}(\dvec|\widehat{\dvec}_i^m)$\\
               & \qquad and 
                 $\overline{\yvec}_{ij}^m\indep\pi(\yvec|\overline{\dvec}_i^m)$, $j=1,\ldots,J_m$ \\
   {\small 9:} & \quad Calculate $\overline{u}_i^m=\prod_{j=1}^{J_m} u(\overline{\dvec}_i^m,\overline{\yvec}^m_{ij})$ \\
   {\small 10:} & \quad Calculate \\
                & \quad $\alpha_i=\min[1,\{\overline{u}_i^m\,q_{MH}(\widehat{\dvec}_i^m|\overline{\dvec}_i^m)\}/\{\widehat{u}_i^m\,q_{MH}(\overline{\dvec}_i^m|\widehat{\dvec}_i^m)\}]$ \\
   {\small 11:} & \quad \textbf{if} $U(0,1)<\alpha_i$ \textbf{then} $\dvec_i^m=\overline{\dvec}_i^m$ \textbf{else} $\dvec_i^m=\widehat{\dvec}_i^m$ \\
   {\small 12:} & \textbf{end} \\
  \caption{Step $m$ of the Resampling-Markov algorithm by \cite{Amzal2006}}\label{alg:Amzal}
  \end{tabular}
\end{algorithm}

\section{Efficiency improvements to the algorithm}
\label{sec:effimp}
We now describe a new algorithm which also uses a particle-based
approach but is one that is much more straightforward, makes more
efficient use of evaluations of the (expensive) utility function and
also more efficiently identifies near-optimal designs. Essentially the
approach bases particle weights on current estimates of expected
utility and thereby focuses sampling effort around near-optimal
designs.

Consider the general case where we need an optimal $k$-timepoint
design $\dvec=(t_1,\ldots,t_k)$, where the times~$t_i$ lie on a grid
rather than in a continuous interval (as in the \Muller, Amzal and ACE
algorithms). This restriction reflects the practical nature of
experimentation but, of course, near continuous designs may be found
by using a grid with a fine mesh.  The algorithm uses refinements of a
particle distribution to increasingly focus on designs around the
optimal design. Instead of basing the weights on increasing powers of
the utility function, in step~$m$ we focus on designs with (current
estimated) expected utility values in the upper $100\alpha_m$\% of
their distribution. The algorithm targets near optimal designs by
working through a series of steps in which the $\alpha_m$--values
decrease as the step number $m$ increases. In this paper we take
$\alpha_m=2^{-m}$.  The powering-up technique used by
\cite{Muller1999} and \cite{Amzal2006} is their way of focusing on
designs which are near optimal. Although we could use a similar
technique which calculates weights by powering-up current estimates of
the expected utility, we feel that it is more natural and intuitive to
deal directly with the size of the upper tail of the distribution of
the expected utility (over designs).

Suppose the location of a design on the discretised mesh of timepoints
is defined by the coordinate system $\lvec=(\ell_1,\ldots,\ell_k)$ for
$\lvec\in\L$ so that the design space is
$\D=\{\dvec_{\lvec}:~\lvec\in\L\}$. The algorithm works with a
(discrete) $k$-dimensional categorical distribution in which the
design at location $\lvec$ has probability $p(\lvec)=w_{\lvec}$.  This
distribution is initialised to be a discrete uniform $\Cat_k(\wvec^0)$
distribution, with un-normalised weights $w_{\lvec}^0=1$, and mass
function $p(\lvec)\propto w_{\lvec}^0,~\lvec\in\L$. This choice is an
exchangeable one and reflects the inability to choose between designs
initially. The initialisation continues by taking a random sample of
design locations from the $\Cat_k(\wvec^0)$ distribution, simulating
datasets from the prior predictive distribution at these designs, and
then determining the utility function at these design/data choices.
These utility calculations are then used to initialise the estimate of
the expected utility
$\hat{u}(\dvec_{\lvec})=\text{mean}(u_{\lvec_k}:~\lvec_k=\lvec)$ and
the number of utility calculations contributing to these means
$n_{\lvec}=\#(\lvec_k=\lvec)$.  Note that any designs not visited
during the initialisation are given zero expected utility. These
estimates of expected utilities are then used to construct the
particle distribution for the first $m=1$ step, $\Cat_k(\wvec^1)$, by
taking $\wvec^1_{\lvec}=\hat{u}(\dvec_{\lvec})$. Note that this choice
is only sensible if all utility estimates are positive.  However, if
this is not the case, as is likely when using say the logged
generalised precision utility function, then we have found using
un-normalised weights $w_{\lvec}= u_{\lvec}-\min_{\lvec'\in\L}
u_{\lvec'}$ works well.

The algorithm then goes through a sequence of steps~$m=1,2,.\ldots,M$
which operate in a similar way to the initialisation.  One potential
issue is that, given the size of the design space, it is possible that
many near-optimal designs are not selected and so it is not prudent to
give these designs zero weight in the particle distribution of the
next step. We circumvent this issue by perturbing the sampled
locations using a distribution~$q$ which, for example, might move a
design to one of its $2^k$ neighbouring locations, and thereby reach
local near-optimal designs. More generally we can use a random walk
proposal such as $q(\dvec_{\lvec^*}|\dvec_{\lvec})=\prod_{i=1}^k
q_i(\ell_i^*|\ell_i)$ over design locations, where each
$q_i(\ell_i^*|\ell_i)$ is a symmetric univariate random walk proposal,
to tailor the size of the local search and help to reach and stay
fairly close to near-optimal designs. One proposal we have found to
work well is to take $\ell_i^*=\ell_i+\nu_i$ where the $\nu_i$ are
independent and have a distribution which is the difference between
two independent Poisson random variables, each with mean~$\lambda$.
Utilities are then calculated by first simulating datasets from the
prior predictive distribution at these design locations and these
utilities used to update the particle weights $\wvec^m$ to contain the
(estimated) expected utility for those designs~$\dvec_{\lvec}$ in the
top $100\alpha_m$\% of the expected utility distribution, with
$\wvec^m_{\lvec}=0$ for all other designs. In the final step, we have
found it useful to increase the accuracy of estimated expected utility
at designs currently thought to be near-optimal, that is, at
designs~$\dvec_{\lvec}$ with $w^M_{\lvec}\neq 0$, through more utility
evaluations. In other words, the final step proceeds essentially the
same as in the previous steps but without any perturbation around the
selected designs.  Finally, after completing all $M$ steps, we take
the optimal design as $\dvec^*=\dvec_{\lvec^*}$, where
$\lvec^*=\arg\max_{\lvec} w_{\lvec}^M$, or conduct a final more
intensive search around this putative optimal design.

\begin{algorithm}[t]
 \begin{tabular}{@{}ll@{}}
   {\small 1:} & \textbf{for} $i=1$ to $N_m$ \textbf{do (in parallel)}\\
   {\small 2:} & \qquad Update un-normalised weights $\wvec^m$ to contain\\
               & \qquad\qquad top $100\alpha_m$\% values of
                  $\{\hat{u}(\dvec_{\lvec}):n_{\lvec}>0\}$ \\
   {\small 3:} & \qquad Simulate $\lvec^*\sim\Cat(\wvec^m)$\\
   {\small 4:} & \qquad Simulate $\lvec\sim q(\lvec|\lvec^*)$ and $\yvec\sim\pi(\yvec|\dvec_{\lvec})$\\
   {\small 5:} & \qquad Calculate $u(\dvec_{\lvec},\yvec)$\\
   {\small 6:} & \qquad Update expected utility\\
               & \qquad\qquad $\hat u(\dvec_{\lvec})\gets\{u(\dvec_{\lvec},\yvec)+n_{\lvec}\hat{u}(\dvec_{\lvec})\}/\{n_{\lvec}+1\}$ \\
   {\small 7:} & \qquad Update count $n_{\lvec}\gets n_{\lvec}+1$ \\
   {\small 8:} & \textbf{end} \\
   {\small 9:} & $\wvec^{m+1}=\wvec^m$\\
  \caption{Step $m$ of new algorithm.}\label{alg:new}
  \end{tabular}
\end{algorithm}

The \Muller, Amzal and ACE algorithms do not make use of utility
calculations made at design locations in previous steps. Thus a simple
but productive efficiency improvement can be made by using all utility
calculations made in previous steps of the algorithm. This is easily
done by keeping a running average of the utility calculations at each
design location.  Additional improvements can be made by using all
computer cores available to the user. For example, if $C$ cores are
available, there is little additional time penalty in calculating $C$
utilities $u(\dvec_{\lvec_k},\yvec_k)$ rather than just one. Finally,
another efficiency gain may be achieved by using a different run
length~$N_m$ in each step, with the earlier steps perhaps using longer
runs. However, the rate at which the $N_m$ decrease should depend on
the extent to which reducing $\alpha_m$ identifies a clear optimal
design. A summary of the new algorithm is given in
Algorithm~\ref{alg:new}. Although updating the weights could be left
until the next step of the algorithm, we have found it beneficial to
update the current weights~$\wvec^m$ regularly, leading to line~2
being within the main loop.

In general, the initial uniform $Cat_k(\wvec^0)$ distribution over
design locations will work reasonably well so long as the number of
possible designs $|\D|$ is not too large. However, if $|\D|$ is large,
for example when using a fairly fine time-grid and seeking a design
with a moderate number of timepoints, making sure that the algorithm
visits all near-optimal designs in step~1 could become problematic.
That said, we have found that expanding the reach of the local random
walk proposals deals with this issue quite well, though this
inevitably leads to needing a larger number of iterations~$N_m$.

\section{Examples}
We demonstrate the efficiency of our method by determining optimal designs of
different sizes in four scenarios. We begin by studying the death model
considered by \citet{CookGG2008} and \citet{DrovandiP2013}. This simple model
has a tractable likelihood and so calculation of the posterior variance, and
hence the utility function, is straightforward. We consider in detail the case
of determining an optimal single timepoint and compare the accuracy of our new
method with those of other popular methods. We then look at finding an optimal
two timepoint design for an oscillatory system typical of those commonly found
when modelling biological systems. The oscillations in this system induce
multiple modes in the expected utility and this complicates the search for
optimal designs. We compare the performance of our algorithm with the ACE
algorithm in determining this optimal two timepoint design. We then compare
performances in finding 15-dimensional optimal designs using a toy utility
function which has features typical of those in real design problems. Finally we
consider optimal design for a more complex stochastic model of aphid growth
\citep{MatisKMS2007}. Here we calculate the posterior variance using a moment
closure approximation to the stochastic model and determine optimal designs of
different sizes using multi-core parallel computing, enabling a six-fold
speed-up.

In the following examples, our algorithm uses threshold
$\alpha_m=2^{-m}$ in step~$m$ and (except where stated otherwise)
spreads the number of utility evaluations equally between the
initialisation and the steps, that is, uses $N_m=N/(M+1)$ particles in
the initialisation and in each step. Also selected designs are
perturbed by (independent) random numbers of grid points (in each
dimension), each calculated as the difference between two independent
Poisson random variates with mean~$\lambda=4$ (except where stated
otherwise).  Runs of the ACE algorithm are made using the
\texttt{acebayes} package with (default) parameters close to
$B=(200,19)$, $N_1=20$ and $N_2=0$ (suggested by the authors) -- the
actual values used were modified slightly to match the computational
budget in each example. The INSH algorithm depends on many more
parameters and we use those provided by the authors and found in their
R~code at \verb#https://github.com/csgillespie/expt_design#.

\subsection{Death model}
\label{sec:death}
\citet{CookGG2008} describe a death model in which the size $Y(t)$ of
the population at time~$t$ obeys the probabilistic law: in the small
time period $(t,t+\delta t]$, the probability of a death is
$Pr\{Y(t+\delta t)=i-1|Y(t)=y(t)\}=\beta y(t)\,\delta t+o(\delta t)$,
otherwise no death occurs. If the population is initialised with size
$n=Y(t_0=0)$ and then observed at times $t_1,\ldots,t_k$, the
likelihood is formed from terms $Y(t_k)|Y(t_{k-1})=y_{k-1}\sim
Bin\{y_{k-1},\exp[-\beta(t_k-t_{k-1})]\}$.

Suppose interest lies in determining the optimal $k$-timepoint design
$\dvec=(t_1,\ldots,t_k)$ using the posterior precision of~$\beta$ as
our utility function, that is, $u(\dvec,\yvec)=1/\Var(\beta|\yvec)$.
One considerable benefit of studying this model is that values of the
expected utility can be calculated and there is no need to employ a
stochastic algorithm such as an MCMC algorithm or importance sampling.
It is fairly quick to calculate
$\K_i(\yvec)=\int\beta^i\pi(\yvec|\beta)\pi(\beta)d\beta$, $i=0,1,2$
over all possible $\yvec=(y_{t_1},\ldots,y_{t_k})$ using the GSL
library \citep{GalassiDTGJABR1996}. Thus we can calculate the expected
utility over all possible designs~$\dvec$ using
\begin{align}
u(\dvec)&=\sum_{y_{t_1}\geq\cdots\geq y_{t_k}=0}^n\pi(\yvec)u(\dvec,\yvec)\nonumber\\
&=\sum_{y_{t_1}\geq \cdots\geq
  y_{t_k}=0}^n\frac{\K_0(\yvec)^3}{\K_2(\yvec)\K_0(\yvec)-\K_1(\yvec)^2},
\label{eq:death}
\end{align}
where $n$ is the initial population size, and thereby determine
the optimal $k$-timepoint design~$\dvec^*$. 
\begin{figure}[t!]
\centering
\includegraphics[width=0.75\textwidth]{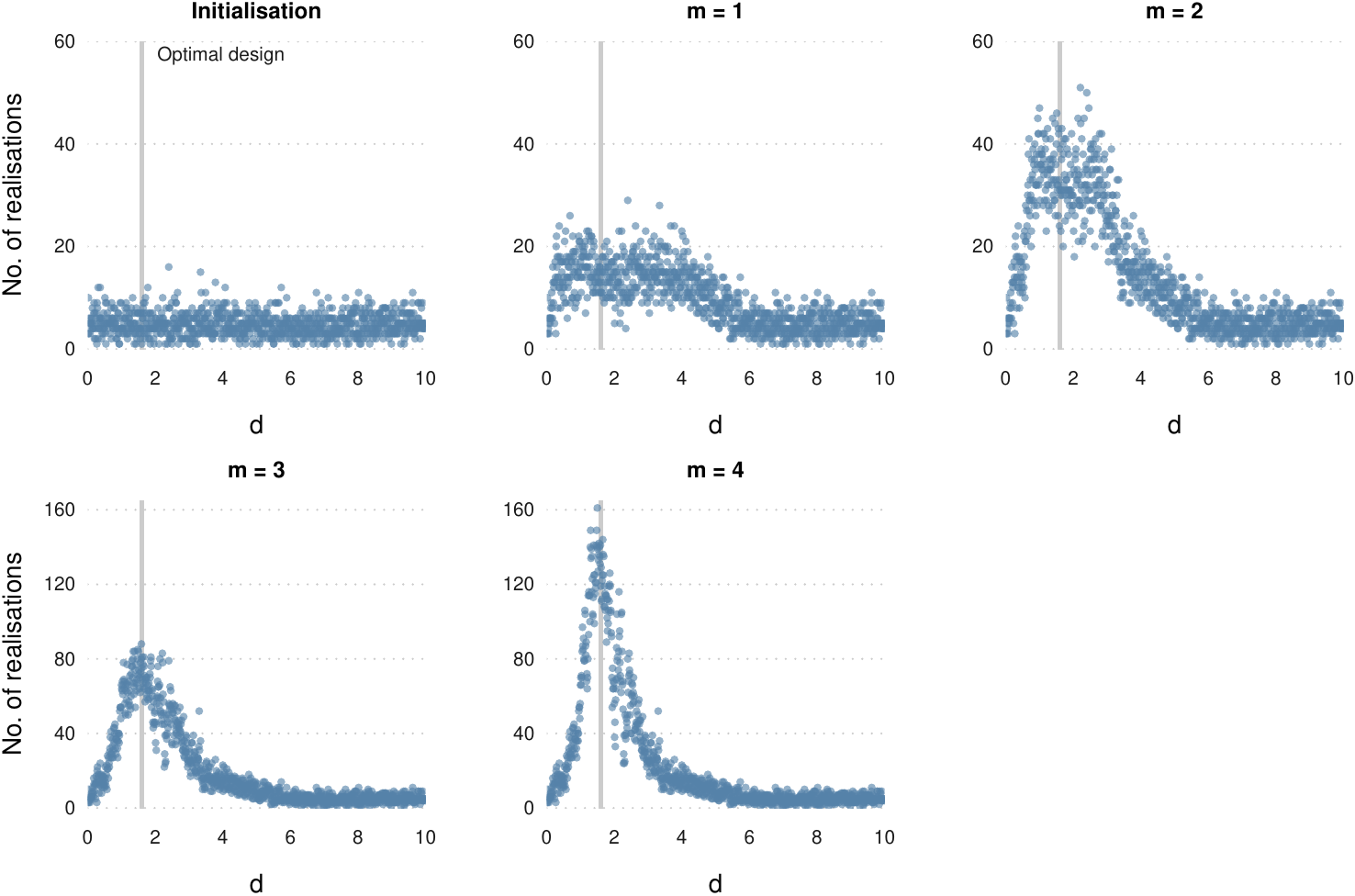}
\caption{Graphs showing the number of realisations $u(d, y_i)$
  contributing to the estimate of $u(d)$ at each $d$ as the number of
  steps $m$ increases. The vertical grey line shows the optimal
  design ($d^*=1.61$).}
\label{F1}
\end{figure}

We follow \citet{DrovandiP2013} by considering an experimental period over
$(0,T=10)$ and restrict design timepoints to be on the grid $t=0.01(0.01)10$.
The initial population size is fixed to be $n=50$. We also take their log-normal
$\LN(-0.005,0.01)$ prior distribution for~$\beta$ and focus on determining the
optimal single $(k=1)$ timepoint design. We also use their utility
function~$\eqref{eq:genprec}$. This problem is sufficiently simple that it is
possible to calculate the expected utility for each possible single timepoint
design and determine that the optimal single timepoint design is
$\dvec^*=t_1^*=1.61$.

Fig.~\ref{F1} shows how the new algorithm
increasingly focuses on getting ever more accurate estimates of near
optimal expected utilities (by averaging over more realisations from
the prior predictive distribution) over the initialisation and steps
$m=1,2,3,4$ of the algorithm, with $\alpha_m=2^{-m}$.  The plot for
the initialisation shows the (near) uniform coverage over all possible
timepoints and then, as $m$~increases, more and more realisations are
simulated at near-optimal timepoints. After the $m=4$ step, the
estimate of expected utility at $t_1^*=1.61$ is an average over $160$
realisations of $u(t_1^*=1.61, y)$.

\begin{figure}[!t]
\centering
\includegraphics[width=0.75\textwidth]{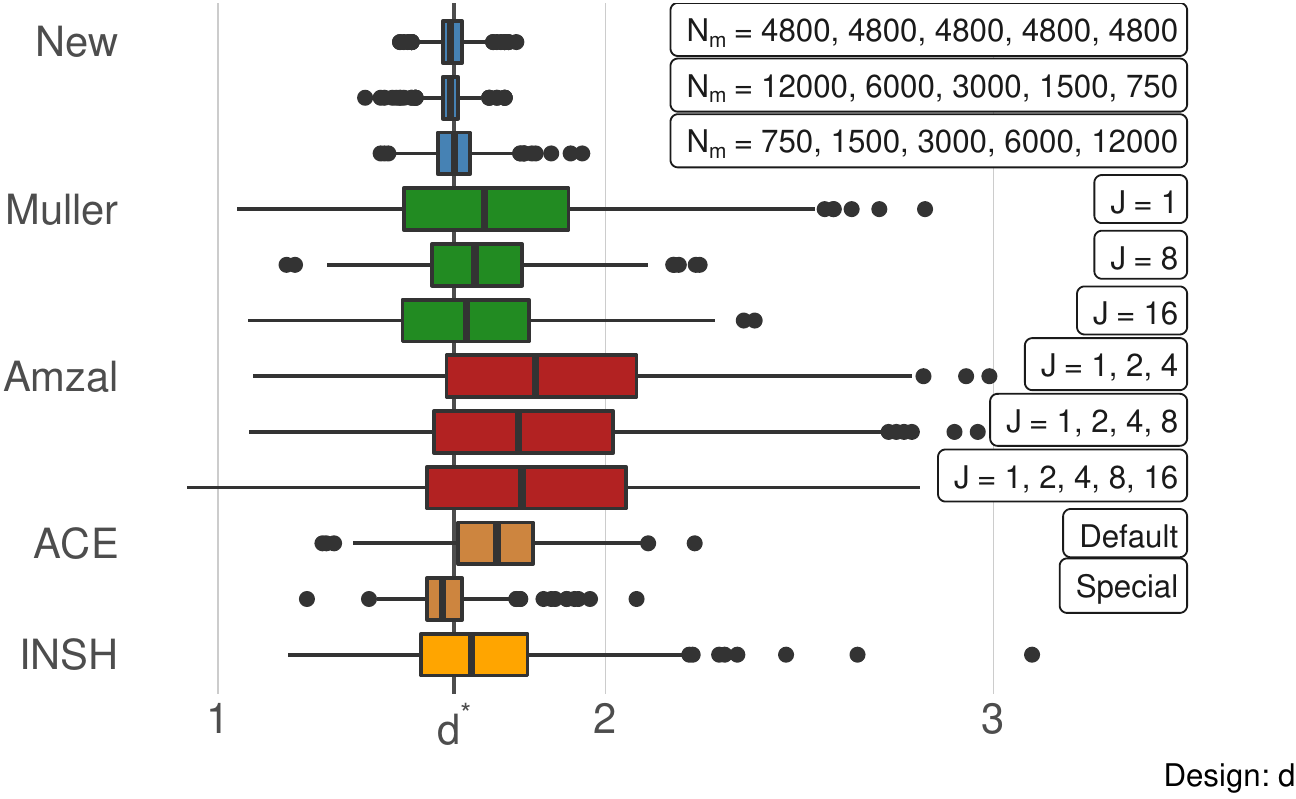}
\caption{Box plots of the optimal designs returned by each algorithm
  from 500 independent runs. The correct optimal design is $d^*=1.61$.
  The different cases are (a) New algorithm with $\alpha_m=2^{-m}$ for
  (i) $N_m=4800, 4800, 4800, 4800, 4800$; (ii) $N_m=12000, 6000, 3000,
  1500, 750$; (iii) $N_m=750, 1500, 3000, 6000, 12000$. (b) \Muller
  algorithm for $J=1,8,16$. (c) Amzal algorithm for (i) $N_m=2400$,
  $J_m=1,2,4$; (ii) $N_m=1090$, $J_m=1,2,4,8$; (iii) $N_m = 522$,
  $J_m=1,2,4,8,16$. (d) Approximate coordinate exchange (ACE)
  algorithm using default and special settings. (e) Induced
  natural selection heuristic (INSH) algorithm.}
\label{F2}
\end{figure}

Fig.~\ref{F2} gives a comparison of the performance of the
\Muller, Amzal, ACE and our new algorithms by showing the sampling distribution
of the optimal designs they return over $500$ independent runs of each
algorithm. Note that we give results for two sets of ACE parameters, one being
the default choice and the other a special choice for this model given to us by
the authors of ACE. The actual values used can be found in our code at
\verb#https://github.com/csgillespie/expt_design#. \\
Each run of each algorithm uses the same computational budget of $24$K utility
evaluations and takes approximately the same run time. Also each algorithm was
run a single CPU core. However, as mentioned previously, if more cores are
available then our new algorithm scales trivially with the number of cores.

The top three boxplots in Fig.~\ref{F2} summarise
the results for the new algorithm but with different a breakdown of
the $24$K utility evaluations in the initialisation and steps
$m=1,2,3,4$. Overall the results show that the algorithm is fairly
insensitive to the number ($N_m$) of evaluations in each step, though
having the $N_m$ decreasing in~$m$ appears to work best.  Also the
boxplots are tightly centered around the optimal design
($t_1^*=1.61$). The next three boxplots are for the \Muller algorithm
with $J=1,8,16$. It is clear that the spread of the \Muller solutions
is much larger than for the new algorithm.  Another issue is that
choosing an appropriate value for~$J$ would require tuning and
therefore additional utility evaluations.  The next three boxplots are
for the Amzal algorithm using different powering up schemes with
different numbers of steps (but still a total of $24$K utility
evaluations). Given the sophistication of the Amzal scheme it is
surprising to see that its performance is similar to the \Muller
algorithm with $J=1$. Inspection of the Amzal algorithm reveals that
this is mainly due to the number of utility evaluations needed within
the main loop (Algorithm~\ref{alg:Amzal}, line~9). The bottom three
boxplots show the results from the ACE and INSH algorithms. The
variation in solutions from ACE is smaller than that of \Muller, Amzal
and INSH, and the INSH solutions are slightly worse than the \Muller
solutions using $J=8$. Perhaps a better overall objective measure of
algorithm performance is their square root mean squared error (RMSE)
about the correct solution $t_1^*=1.61$. The values for these various
implementations (in the order top-bottom in
Fig.~\ref{F2}) is New: 0.04, 0.07, 0.04, \Muller:
0.33, 0.18, 0.24, Amzal: 0.43, 0.41, 0.44, ACE: 0.08, 0.17, INSH: 0.23
and show that the new algorithm can perform between 2.5 and 11 times
more efficiently.  An additional and powerful attribute of the new
algorithm is that it is a much more simple algorithm to implement than
the others and is embarrassingly parallel.

\begin{figure}[t]
\centering
\includegraphics[width=0.48\textwidth]{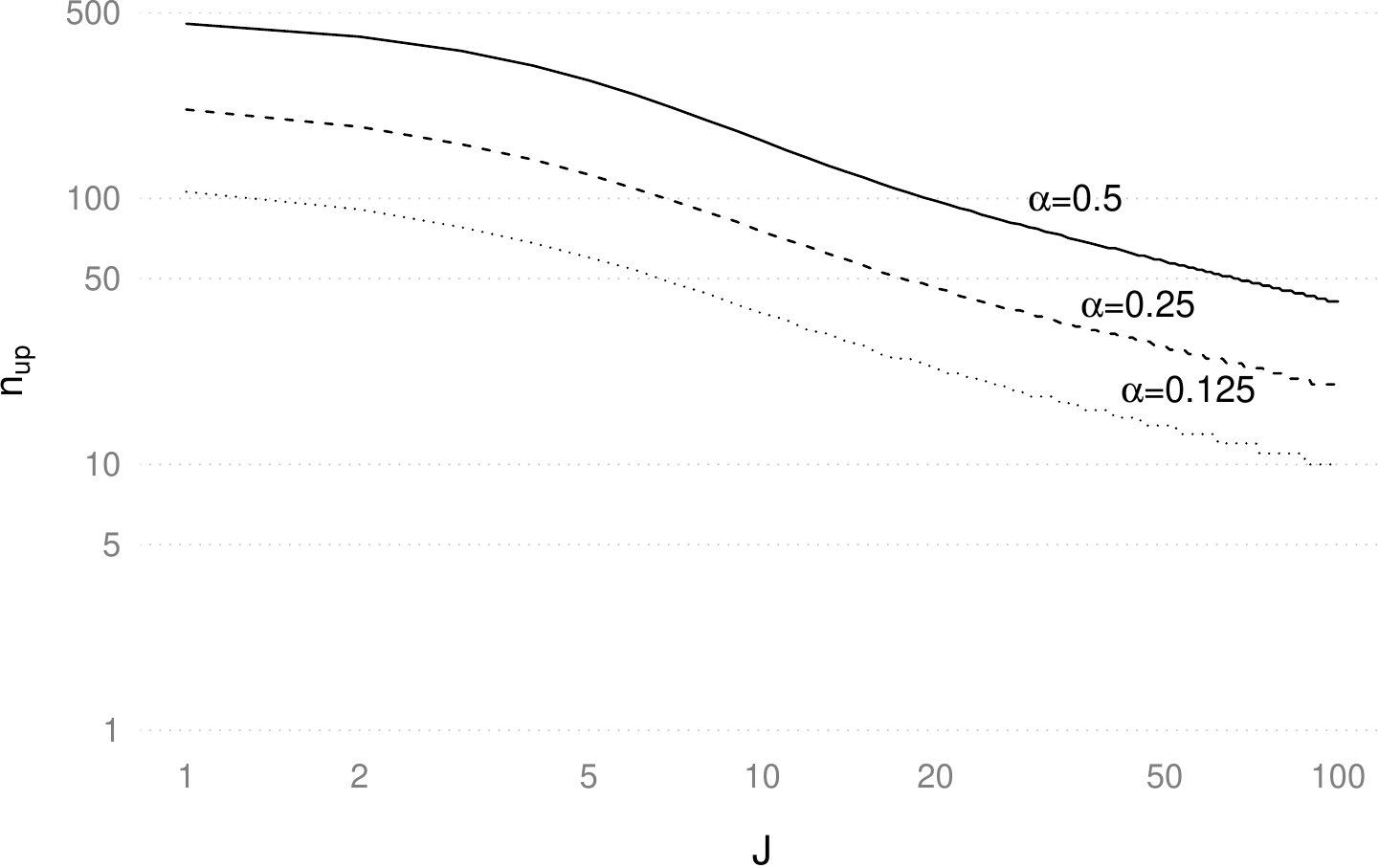}
\caption{Relationship between the level of powering up ($J$) of the
  utility function in the \Muller algorithm and, for the new
  algorithm, the number of designs ($n_{up}$) in the upper tail of the
  expected utility distribution, calculated over all $1000$ single
  timepoint designs for the death model.}\label{F3}
\end{figure}

There is a indirect relationship between decreasing~$\alpha_m$ in the
new algorithm and \Muller's powering-up approach. Consider the
normalised utility values
%\begin{equation}\label{2}
%w_i^J = \frac{u (d_i)}{\sum_{i'=1}^{|\D|} u^J (d_{i'})}
%\end{equation}
$w_i^J=u (d_i)/\sum_{i'=1}^{|\D|} u^J (d_{i'})$, with order statistics
$w_{(1)}^J\leq \cdots\leq w_{(|\D|)}^J$ and empirical distribution
function~$\hat{F}_J(\cdot)$. A measure of the correspondence between
the new algorithm and its \Muller equivalent is the value of~$k$,
where
\begin{equation}\label{3}
  \arg\min_k \hat{F}_J\bigl\{w_{(k)}^J\bigr\}>1-\alpha_m,
\end{equation}
and $\alpha_m$ is the threshold used in step~$m$ of the new algorithm.
The distribution of expected utility becomes increasingly peaked as
$J\rightarrow\infty$, in which case $w_{(|\D|)}^J\to 1$, that is,
$k\rightarrow|\D|$. Also during the $m=1$ step of the new algorithm,
designs are sampled from a distribution with weights proportional to
(an estimate of) $u(d)$ and so this corresponds to the \Muller
algorithm with $J=1$. In the \Muller algorithm, as the expected
utility is powered-up (by increasing $J$) the algorithm preferentially
sample designs near the optimal design, that is, in the upper tail of
the expected utility distribution. Unfortunately it is difficult to
obtain an algebraic understanding of how increasing~$J$ focuses on
designs further into the upper tail of the distribution of expected
utility values, that is, its effect on~$k$. However we can calculate
the (exact) expected utility (for this simple death model) at all
1000~single timepoint designs in the design space using
\eqref{eq:death} and thereby determine values of~$k$ for different
choices of~$J$ for various values of~$\alpha_m$; see Fig.~\ref{F3}.
For example, when $\alpha_m=0.5$ and $J=1$, we obtain $k=454$ as the
sum of the largest $454$ normalised utility values is greater than 0.5
(but that of the largest $453$ is not). We see that $k$ decreases as
$J$ increases (for fixed $\alpha_m$). However for $J>20$ the change in
$k$ drastically slows down.  Furthermore, using values of $J>100$,
results in numerical issues. Also, although the traces in~$k$ for
different~$\alpha_m$ look parallel, they are only roughly parallel
with, for example, $k_{\alpha=0.5}/k_{\alpha=0.25}$ ranging between
2.0 and 2.3. The figure highlights the main issue with the powering up
approach: whilst being an intuitively good idea, in practice the rate
at which this homes in on the optimal design as $J$ increases is
unclear and difficult to predict. Also working with large powers of
utilities often introduces problems of numerical instability.

\subsection{Oscillatory systems}
Solutions to deterministic or stochastic descriptions of biological
systems often display oscillatory behaviour, particularly those
describing homoeostasis maintained by regulatory mechanisms.  Here we
consider a toy model which exhibits this behaviour but also has a
straightforward conjugate Bayesian analysis. The model is one of
damped oscillations where observations follow
\begin{equation}
y_t=\theta e^{-t}\sin 6\pi t + \eps_t,
\label{eq:dampedsoln}
\end{equation}
with $\eps_t\indep\N(0,\sigma^2)$ and $t\in[0,T=1]$. The (unknown)
parameters are the size of the oscillations (described by~$\theta$)
and the level of observational noise~($\sigma$). A key feature of this
model is that multiple cycles can be observed within the observed time
period.

If this observational model is rewritten as $y_t=\theta f(t) + \eps_t$
then it is clear that it is a simple normal regression model. Such
models have a conjugate normal-gamma prior distribution, which in this
case is $\theta|\sigma\sim\N(b,\sigma^2/c)$ and
$\sigma^{-2}\sim\Ga(g,h)$. The posterior distribution after observing
data~$\yvec$ from a design with $k$ timepoints takes the same form
with $\theta|\sigma,\yvec\sim\N(B,\sigma^2/C)$ and
$\sigma^{-2}|\yvec\sim\Ga(G,H)$, where $B=(bc+P)/C$, $C=c+Q$,
$G=g+k/2$,
$H=h+[\sum_{i=1}^k\{y_{t_i}-Pf(t_i)/Q\}^2+b^2c+P^2c/(QC)]/2$ with
$P=\sum_{i=1}^kf(t_i)y_{t_i}$ and $Q=\sum_{i=1}^kf(t_i)^2$. It is
easily shown that $Cov(\theta,\sigma^2|\yvec)=0$ and so the posterior
generalised precision for $(\theta,\sigma^2)$ is $C(G-1)^3(G-2)/H^3$.
Therefore, as $G$ does not depend on data values, we take our utility
function as the logged generalised precision $u(\dvec,\yvec)=\log
C-3\log H$.

\begin{figure}[t]
\centering
\includegraphics[width=0.48\textwidth]{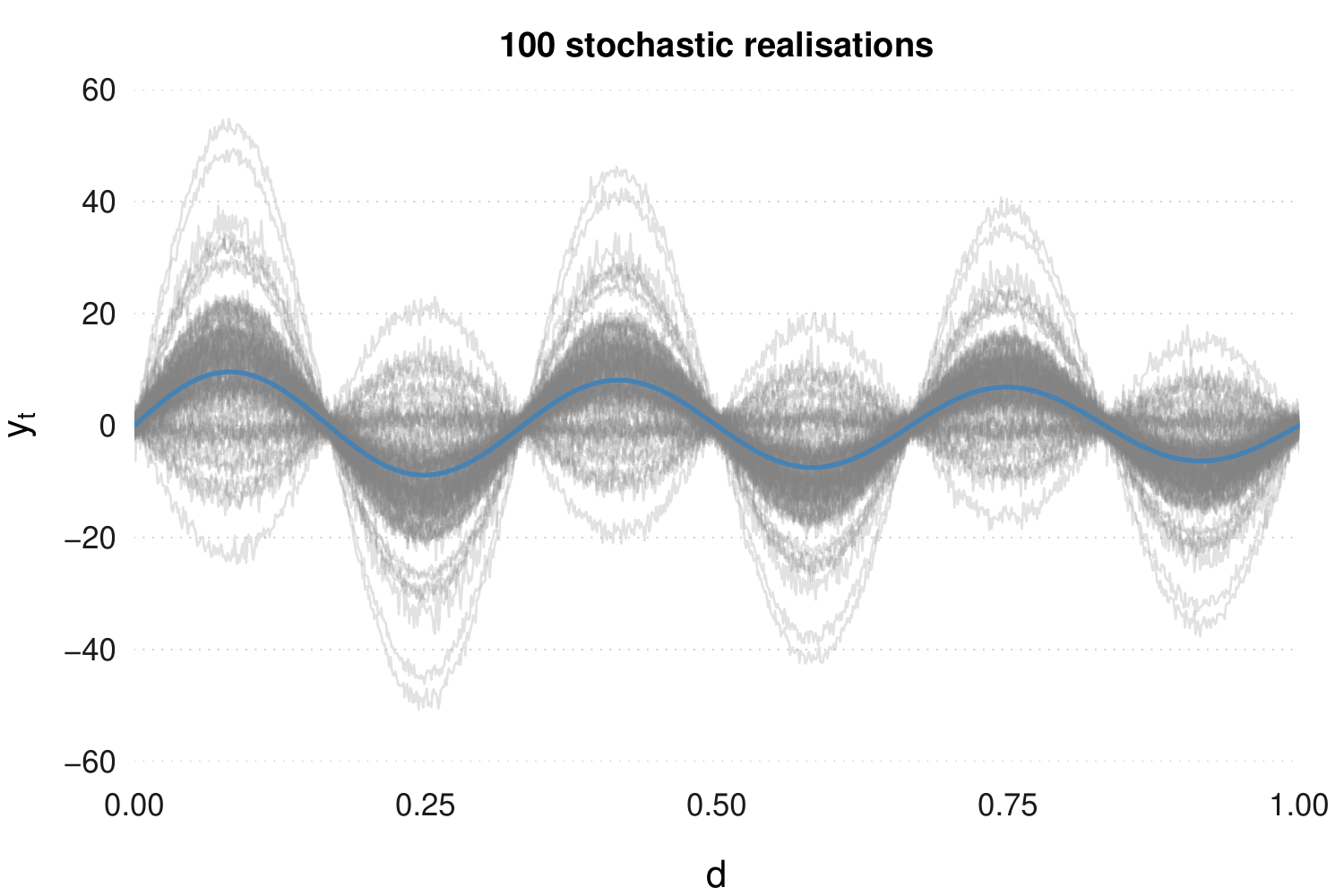}
\caption{100 prior predictive realisations from the oscillatory system model,
  together with the prior predictive mean trace.}\label{F4}
\end{figure}

We now compare the performance of our algorithm with that of ACE in
determining the optimal $d=2$ time-point design. We construct the
prior distribution so that values of $\theta$ and $\sigma$ are
typically around ten and one respectively but fairly uncertain
($b=10$, $c=0.01$, $g=3$ and $h=3$). Figure~\ref{F4} shows typical
realisations from the prior predictive distribution, together with the
prior predictive mean trace. We will again assume a computational
budget of $24$K utility evaluations for each algorithm. For our
algorithm we use $N_m=2.4$K particles in the initialisation and in
each step $m=1,2,\ldots,9$ and search for optimal designs on a grid
$t=0(0.002)1$.  Random walk perturbations were made using $\lambda=4$
except in the final step ($\lambda=0$).

\begin{figure}[t]
\centering
\includegraphics[width=0.48\textwidth]{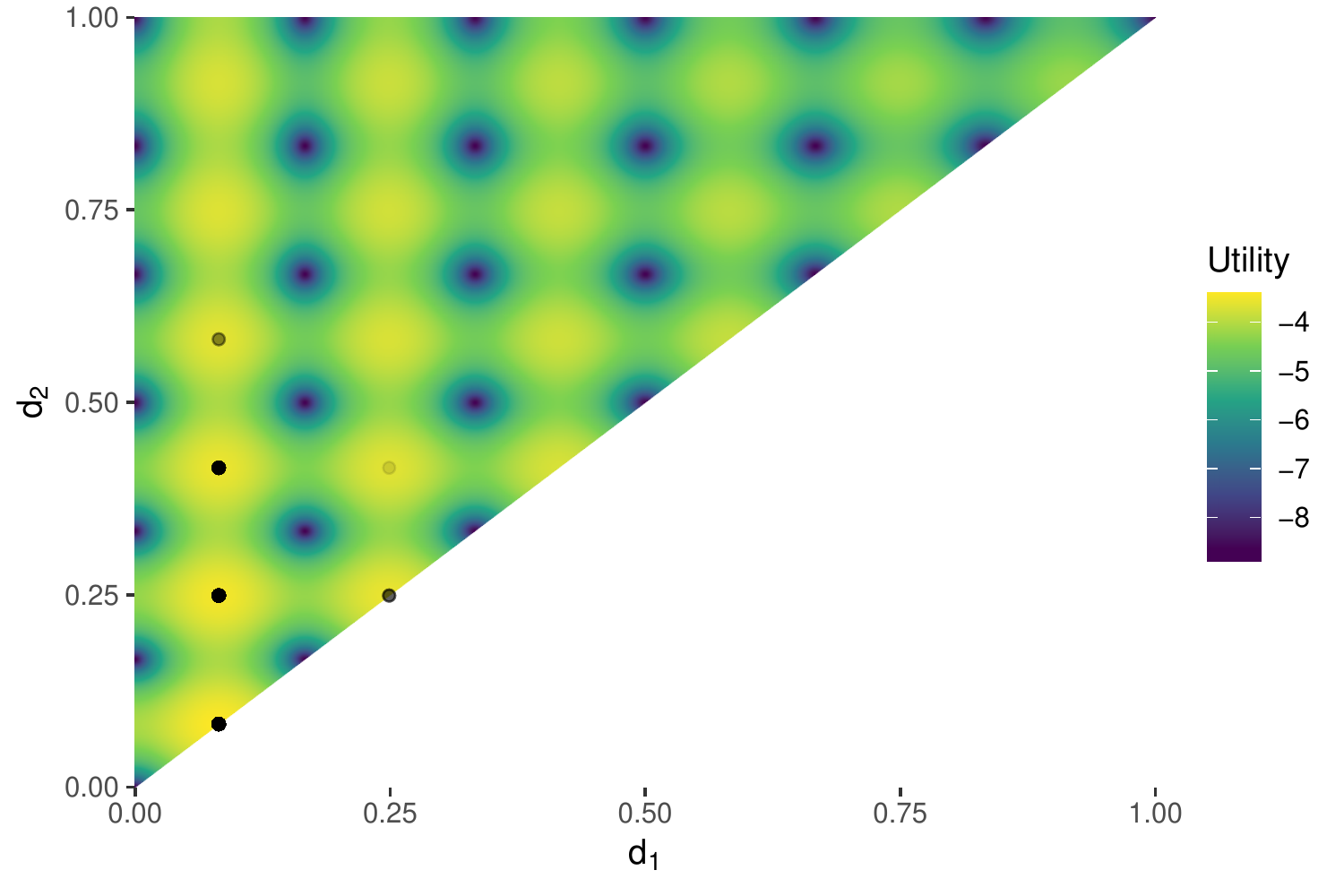}
\includegraphics[width=0.48\textwidth]{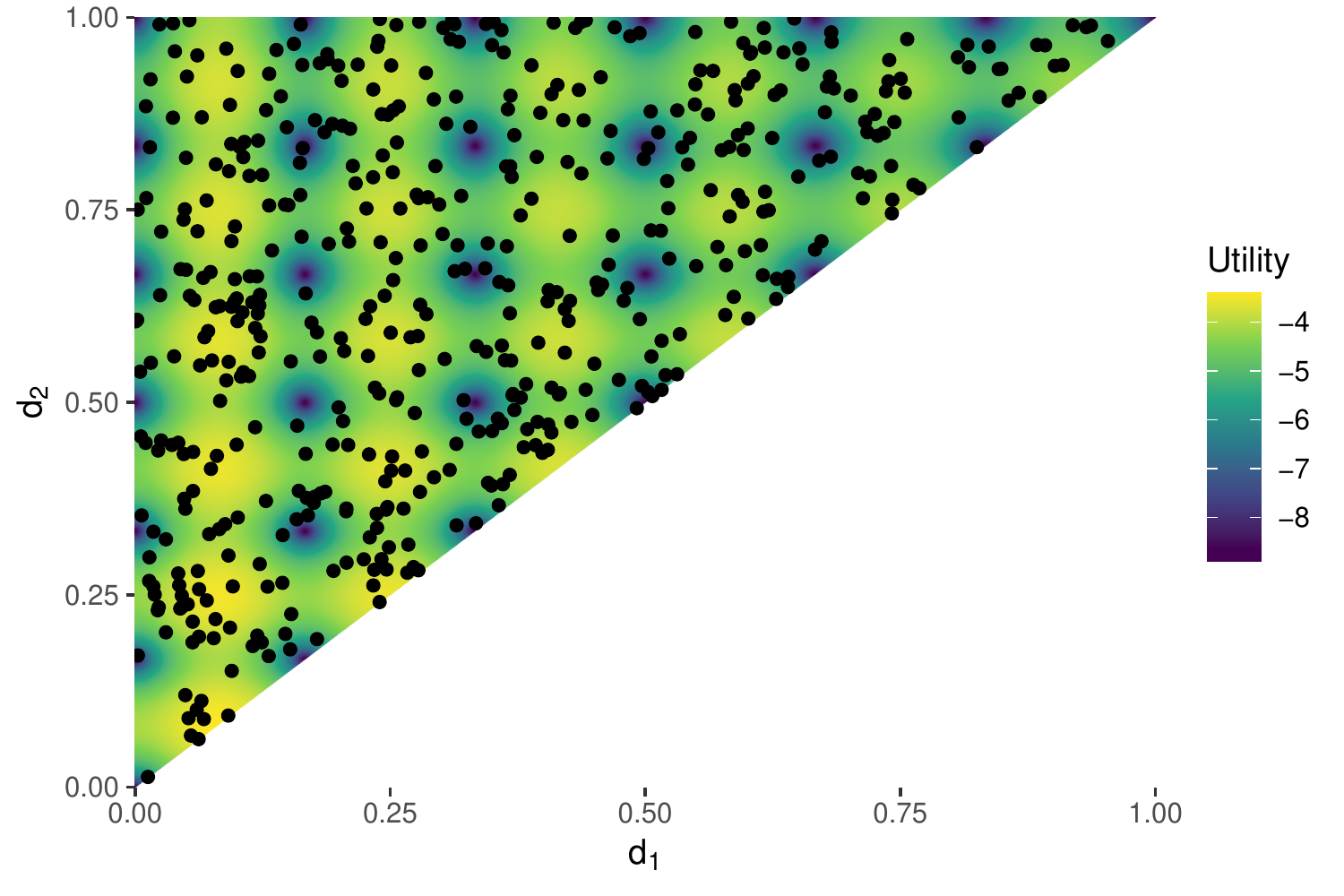}
\caption{500 optimal designs determine by our new algorithm (top) and
  the ACE algorithm (bottom), together with contours of the expected
  utility surface.}\label{F5}
\end{figure}

Figure~\ref{F5} shows the optimal designs obtained from 500 runs of
each algorithm. It also shows the multi-modal nature of the underlying
expected utility surface. Clearly the ACE algorithm gets stuck in
local modes whereas our algorithm hits only local modes with the
correct optimal design (two replicates at time $t=0.082$) occurring on
48\% of occasions. The performance of the ACE algorithm is perhaps not
surprising as it initialises at a random design and works best with
uni-modal expected utility surfaces (for high dimensional designs).
Although ACE is likely to perform much better with more utility
evaluations, we did find that the ACE solutions were similarly
scattered after using $48$K evaluations, whereas our algorithm found
the correct optimal design 96\% of the time.

\subsection{High dimensional designs}

\begin{figure}[t]
\centering
\includegraphics[width=0.48\textwidth]{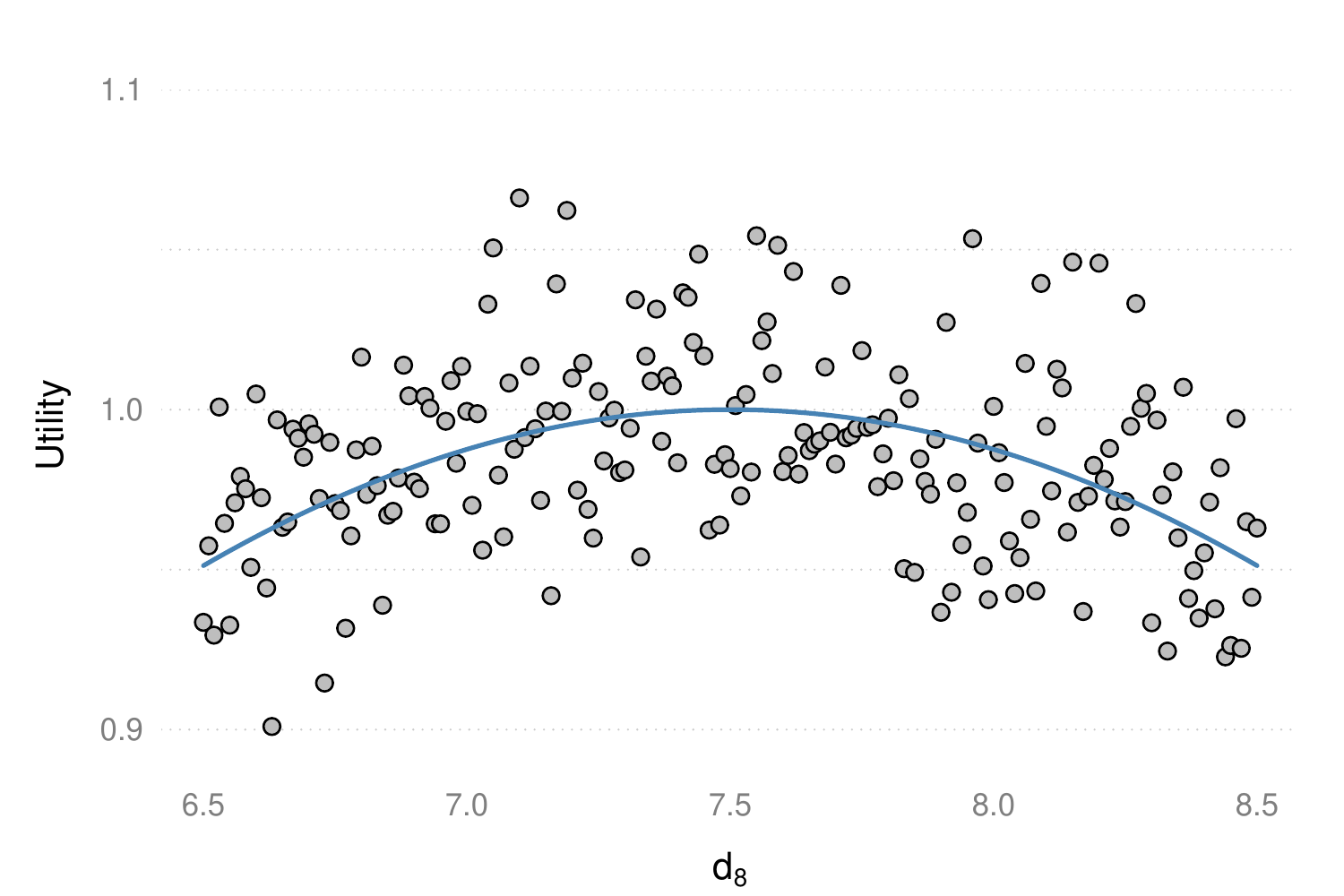}
\caption{The median utility function and its realisations at the
  15-dimensional design with middle time-point $d_8=6.5(0.01)8.5$ and
  other time-points at their optimal choice ($d_i=\mu_i$, $i\neq
  8$).}\label{F6}
\end{figure}

The \Muller and Amzal algorithms are not computationally efficient
when determining high dimensional designs. However, the ACE algorithm
has been designed to solve this problem in a computationally efficient
way. We now compare the performance of the ACE algorithm and our
algorithm in determining a $d=15$ time-point design. As we need to run
the algorithms many times we have chosen to examine the performance of
these algorithms using a utility function which is quick to evaluate
but also has features seen in real problems, namely that it is
unimodel and fairly flat. We base our utility function on a
$15$-dimensional normal density with mean vector
$\muvec=(0.5,1.5,\ldots,14.5)^T$ and covariance
matrix~$\Sigma=10I_{15}$, where $I_{15}$ is the $15\times 15$ identity
matrix.  Specifically we take the utility function for data~$\yvec_i$
to be
\[
u(\dvec,\yvec_i)=\exp\left\{-(\dvec-\muvec)^T(\dvec-\muvec)/20\right\}\,\eps_i
\]
where $\eps_i\indep\LN(0,0.03^2)$. Note that this scaling of the
normal density gives it a maximum value of one.  Figure~\ref{F6}
displays the median utility function and utility realisations for the
design with middle time-point $d_8=6.5(0.01)8.5$ and other time-points
at their optimal choice ($d_i=\mu_i$, $i\neq 8$). The figure shows a
typical view of the median utility function: it is unimodal, fairly
flat and its realisations are quite noisy.

We now look for the optimal 15~time-point design in the observational
period $[0,T=15]$ using a fairly fine grid $t=0(0.01)15$. We compare
the algorithms assuming a computational budget of $360$K utility
evaluations. For our algorithm, after the initialisation we move
through steps $m=1,2,\ldots,14$ with $N_m=24$K particles in each
step using random walk perturbations with $\lambda = 1$, except in the
final step ($\lambda=0$). The algorithms were run on a single CPU and
the computational time taken for our algorithm was around ten times
that for ACE. This is because our algorithm is designed to sacrifice
storage for speed and, in particular, retrieving past results can be
(relatively) time-consuming. However employing ten CPUs in parallel
would give the algorithms a similar duration. Also, for more realistic
models, the utility evaluation would be significantly longer and so
running our algorithm in parallel would be quite beneficial, if not
essential.

The RMSE of 100 algorithm solutions about the correct optimal design
$\dvec^*=\muvec$ is New: 0.01 and ACE: 0.001, with the best expected
utility found in these solutions being New: 0.98 and ACE: 0.9997 to be
judged against the maximum achievable expected utility $u(\dvec^*)=1$.
Clearly ACE out-performs our algorithm in determining the 15-d optimal
design. However our algorithm performs reasonably well. 
       
\subsection{Cotton aphids}

A cotton aphid infestation of a cotton plant can result in many
problems such as leaves that curl and pucker, seedling plants become
stunted and may die, a late season infestation can result in stained
cotton. Also cotton aphids have developed resistance to many chemical
treatments and so can be difficult to treat. Therefore considerable
effort and cost is used in the maintenance of cotton plants.
\citet{MatisKMS2007} have developed a stochastic model of aphid
population growth. \citet{GillespieG2010} give a Bayesian analysis of
this model using data given in \citet{MatisPMMS2008}.  The data
contain aphid counts on twenty randomly chosen leaves in each plot,
for twenty-seven treatment-block combinations. The treatment-blocks
were formed from three three-level factors (nitrogen and irrigation
levels and block). Observations were taken roughly every 7 to 8 days
within a 32 day period.

Let $N(t)$ and $C(t)$ denote the size and cumulative size of the aphid
population respectively at time $t$. \citet{MatisKMS2007} modelled
aphid dynamics using a birth rate of $\lambda N(t)$ and a death rate
of $\mu N(t)C(t)$. Therefore, in a small time period $(t,t+\delta t]$,
  so that at most one event can occur, we have
\begin{align*}
Pr\{N(t+\delta t)&=n(t)+1,C(t+\delta t)=c(t)+1|n(t),c(t)\}=\lambda n(t)\,\delta t+o(\delta t),\\
Pr\{N(t+\delta t) &=n(t)-1,C(t+\delta t)=c(t)|n(t),c(t)\}=\mu n(t)c(t)\,\delta t+o(\delta t),
\end{align*}
and the probability of staying in the same state is one minus the sum
of these probabilities.

\citet{GillespieG2010} analysed these data by making a normal approximation to
the stochastic model using moment closure. This gives transition distributions
\[
(N(t_i),C(t_i))^T|N(t_{i-1}),C(t_{i-1}),\lambda,\mu
\] which
follow~$N_2\{\mvec(t_{i-1}),\vvec(t_{i-1})\}$ distributions, where
$\mvec(t)=(m_1(t)=\E\{N(t)\}$, $m_2(t)=\E\{C(t)\})^T$ and
$\vvec(t)$ is a matrix with diagonal elements
$v_{11}(t)=\Var\{N(t)\}$ and
$v_{22}(t)=\Var\{C(t)\}$, and off-diagonal elements
$v_{12}(t)=\Cov\{N(t),C(t)\}$. Note that the
$(\lambda,\mu)$-dependence of these functions has been suppressed to simplify
the exposition. These functions are determined as solutions of the ODE system
\begin{align}
\frac{dm_1(t)}{dt} &= \lambda m_1(t)- \mu\{m_1(t)m_2(t)+v_{12}(t)\} \label{eq:lna}\\ 
\frac{dm_2(t)}{dt} &= \lambda m_1(t) \notag
\end{align}
\begin{align}
\frac{dv_{11}(t)}{dt}&=\mu[v_{12}(t)-2m_1(t)v_{12}(t)-2\kappa_{21} +m_2(t)\{m_1(t)-2 v_{11}(t)\}]\\
&\qquad\qquad+\lambda\{m_1(t)+2v_{11}(t)\} \notag\\
\frac{dv_{12}(t)}{dt}&=\lambda\{m_1(t)+v_{11}(t)+v_{12}(t)\} -\mu\{m_1(t)v_{22}(t)+m_2(t)v_{12}(t)\} \notag\\
\frac{dv_{22}(t)}{dt}&=\lambda\{m_1(t)+2v_{12}(t)\},\notag
\end{align}
with initial conditions $m_1(0)=n_0$, $m_2(0)=c_0$ and
$v_{11}(0)=v_{12}(0)=v_{22}(t)=0$. This system can be solved
numerically using standard ODE solvers. Note that this approximation
is the same as that when using the linear noise approximation if the
term $v_{12}(t)$ in $\eqref{eq:lna}$ is ignored.

We now construct optimal designs assuming a similar treatment-block
pattern and observation period. The aim is to optimise the posterior
generalised precision for the rate parameters
$\thetavec=(\lambda,\mu)$.  \citet{GillespieG2010} found only small
differences in the rate parameters for the various treatments and
blocks and so we base our prior distribution on that of the posterior
distribution for the base treatment and block rates, however we
inflate prior uncertainty by increasing the variances by a factor of
ten, giving
\[
\begin{pmatrix}
\lambda\\
\mu 
\end{pmatrix}
\sim N_2 \left\{
\begin{pmatrix}
0.246\\
0.000134
\end{pmatrix},
\begin{pmatrix}
0.0079^2 & 5.8 \times 10^{-8}\\
5.8 \times 10^{-8} & 0.00002^2
\end{pmatrix}\right\}.
\]
Also to simplify the analysis, we assume known initial aphid levels
$(n_0=c_0=28)$ as there was very little variability in these
quantities in the \citet{MatisPMMS2008} dataset.

\begin{figure}[t]
\centering
\includegraphics[width=0.65\textwidth]{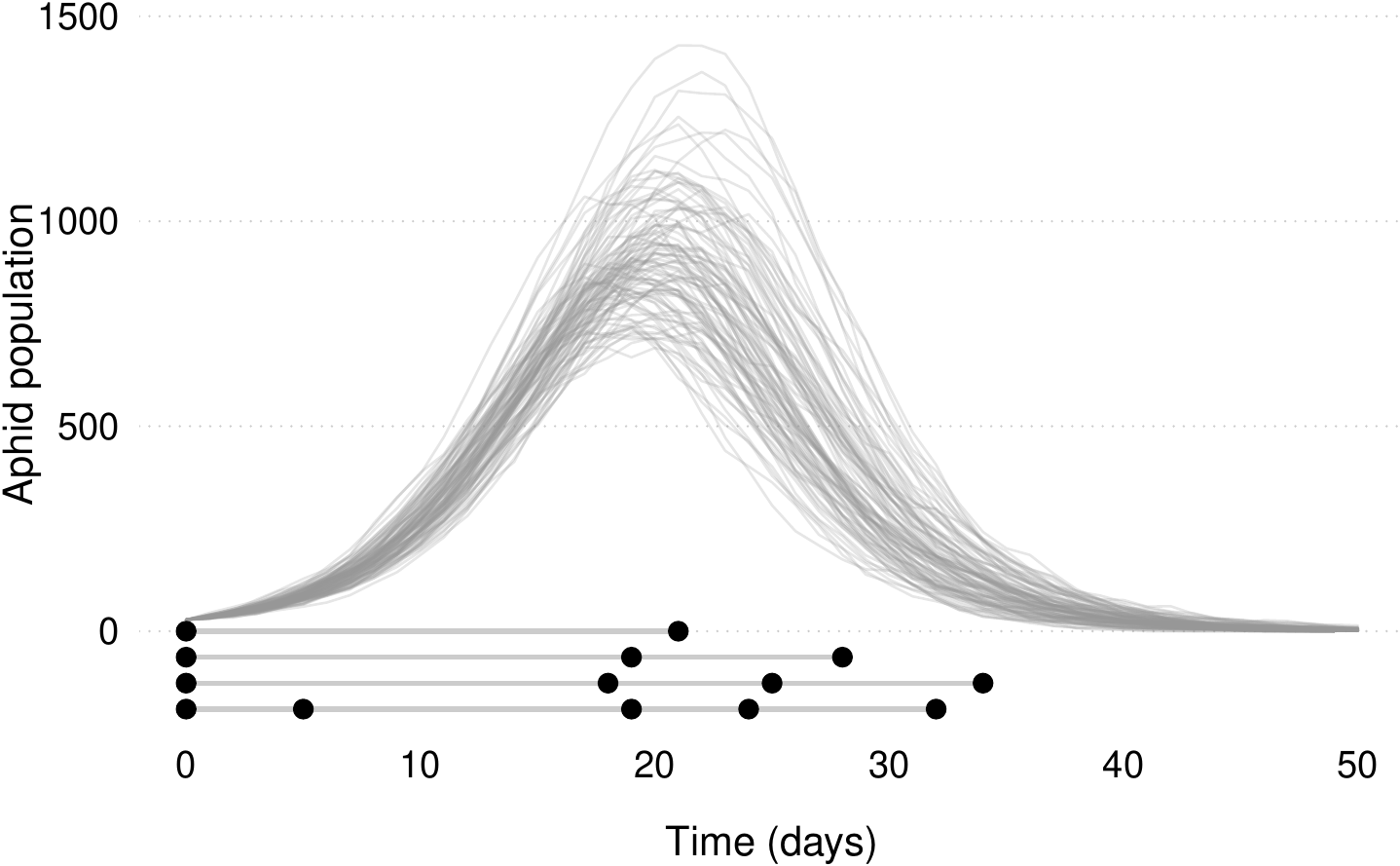}
\caption{100 prior predictive realisations together with the optimal
  one, two, three and four timepoint designs.}\label{F7}
\end{figure}

One hundred simulations from the prior predictive are shown in
Fig.~\ref{F7}.  All simulations ultimately end in the extinction of
the aphid population, since the cumulative aphid population is
increasing with each birth, resulting in $\mu n(t) c(t) > \lambda
n(t)$ as $t$ increases. All simulations show a peak in aphid
population around 20--25 days.

The posterior density for the rate parameters when using a design with
$k$ timepoints yielding data $\yvec$ is
\[
\pi(\lambda,\mu|\yvec) \propto 
\pi(\lambda,\mu)\prod_{i=1}^k \phi_2\{n(t_i),c(t_i)|\mvec(t_{i-1}),\vvec(t_{i-1})\},
\]
where $\phi_2(\cdot,\cdot|\mvec,\vvec)$ is the $N_2(\mvec,\vvec)$
density.  Realisations from this posterior were obtained using an MCMC
scheme with a bivariate normal random walk proposal, centred at the
current value and a covariance matrix with standard deviations
$0.0009$ and $0.000004$, and correlation equal to the prior
correlation. We seek optimal $k=1, 2, 3, 4$ timepoint designs within a
$T=49$ day period. Note that this period is slightly longer than that
used in the original experiment so we can investigate whether the
experiment was stopped too early. Thus the design timepoints are in
units of 24 hours (called days) after the start of the new experiment
(day zero).  Again each posterior distribution was determined by
initialising the chain at the parameter values used to simulate the
responses~$\yvec$ and we found that very little burn-in was necessary.
Each time the algorithm was run for 10K iterations, and typically took
no more than two~cpu seconds on an Intel Core i7-6700 CPU.

\begin{figure}[t]
\centering
\includegraphics[width=0.9\textwidth]{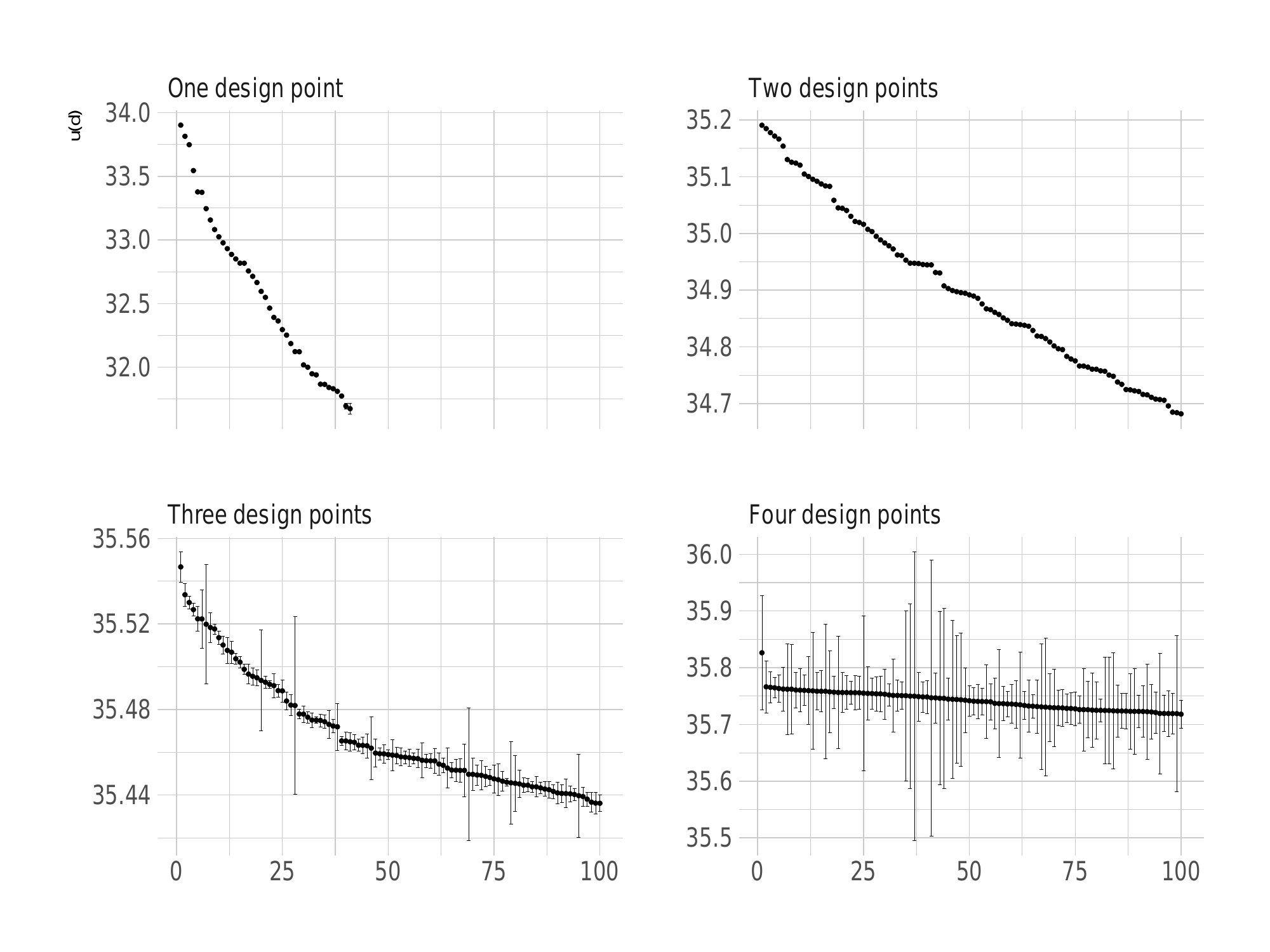}
\caption{Estimated expected utilities (and central 95\% intervals) for
  the top 100 designs with 1, 2, 3 and 4 timepoints.}\label{F8}
\end{figure}

Again we determine the optimal design using the predictive precision
utility function $\eqref{eq:genprec}$. We look for one, two, three and
four--timepoint optimal designs by moving through
steps~$m=1,2,\ldots,8$ after initialisation. We use a computational
budget of $432$K utility evaluations ($48$K in each of the
initialisation and each step) and use random perturbations with
$\lambda=4$ except in the final step ($\lambda=0$). 

Fig.~\ref{F8} shows the top 100 designs (by estimated expected
utility) with 1, 2, 3, and 4 timepoints together with (rough) 95\%
confidence bounds calculated assuming asymptotic posterior normality.
Note that there are only 50 possible single timepoint designs. One of
the benefits of using our algorithm is that it is straightforward to
keep track of the precision of the current expected utility estimates
by also keeping a running total of the $u(\dvec,\yvec_i)^2$. The
figure shows that the confidence bounds for the one and two time point
designs are very small. The intervals for the three point design are
also relatively narrow. However, the optimal design is not clear and
further runs will be needed to reduce uncertainty on the expected
utility estimates.  There is much more uncertainty in expected utility
estimates for the four design point problem and clearly many more
simulations will be needed to pick out the optimal design. All that
said, the differences in expected utility for all top 100 designs (of
a given size) are quite small.

\begin{figure}[t]
\centering
\includegraphics[width=0.6\textwidth]{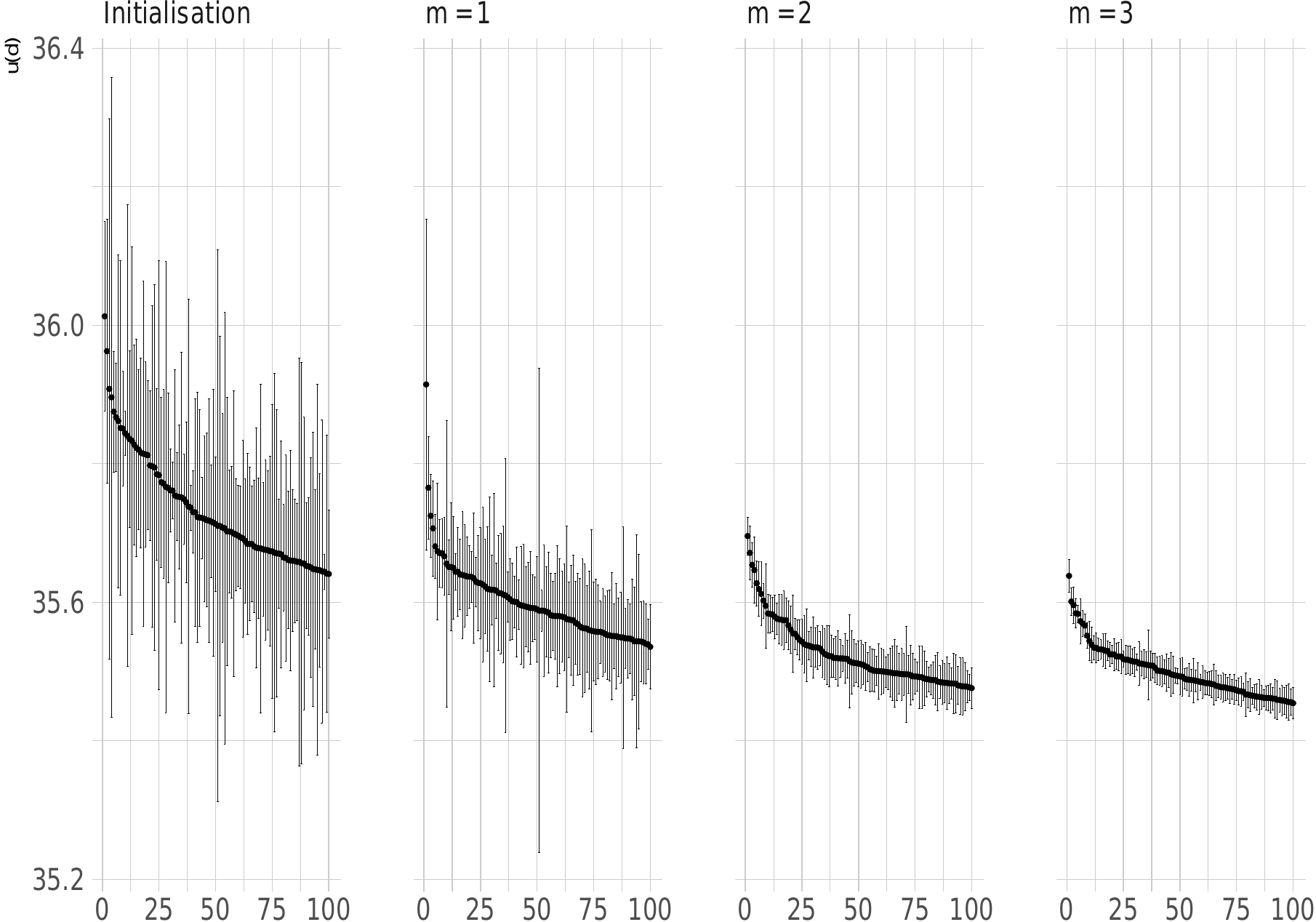}
\caption{Estimated expected utilities (with central 95\% confidence intervals)
  for the top~100 three timepoint designs for the aphid model after the
  initialisation and each step.}\label{F9}
\end{figure}

One advantage of our new algorithm (over \Muller and Amzal) is that it
is trivial to extend a search for an optimal design so long as the
particle weights are retained from the previous run. A further
advantage is that any extension to the optimal search run is not
restricted to being on the same computer resource (desktop, cluster,
cloud) as the initial run. This can be advantageous when determining
the optimal design for a complex model which has a time consuming
utility calculation as decisions on numbers of particles and their
location can be amended easily (by the user) at each step of the
algorithm. Also inspection of the (estimated) expected utilities can
be examined at each step of the algorithm to determine how many steps
are needed before the location of the optimal design is clear. For
example, Fig.~\ref{F9} shows the top 100 three timepoint designs for
the initialsation and steps~$m=1,2,3$ and shows that uncertainty
around the top designs is sufficiently small that there is no need to
extend the algorithm to another step to determine the top five designs
in contention for being optimal design.

\section{Conclusion}

The search for Bayes optimal designs is time consuming as it often
requires the evaluation of a non-trivial utility function at very many
datasets simulated from the prior predictive distribution. In this
paper we have focused on utility functions that depend on the
generalised posterior precision, and generally this can be fairly slow
to evaluate if, for example, it is estimated by determining the
parameter posterior distribution via MCMC. The search is further
complicated because, in general, the expected utility function is
fairly flat. In contrast to other methods, we use a stepwise approach
to focus in on designs in the upper tail of the distribution of
expected utility, as we believe this measure to be more intuitive than
criteria used in other methods.

This new algorithm is generally much more efficient in determining the optimal
design in stochastic process models than others available in the literature,
though the ACE algorithm performs particularly well in determining large designs
when the expected utility function is unimodal. The algorithm also out performs
others when the expected utility function is multi-modal and the computational
budget is fairly limited. It uses a particle representation over design space to
focus increasingly on regions of high utility. There are no issues of
convergence in the scheme (as there are in, for example, the \Muller scheme) and
the output facilitates a simple comparison of near-optimal designs via their
(estimated) expected utility allowing for uncertainty in the estimates.

The new algorithm is initialised using a variation of the general
scheme.  However, it could be initialised in a variety of ways. For
example, a (space filling) maximin Latin hypercube design could be
used, as is typical in the calibration of stochastic emulators
\citep{BaggaleyBGSS12}. Alternatively the algorithm could be first run
using a delta approximation to the expected utility function to find
designs maximising $u\{\dvec,E(\yvec)\}$ and then determining the
initial weights using the evaluations of this approximate expected
utility function. The algorithm is embarrassingly parallel and ideally
suited to running on multi-core computers. The search for the optimal
design is easily interrupted and restarted, allowing for the search to
be monitored and additional compute resource to be added.

\bibliographystyle{Chicago}

\bibliography{full,paper}

\end{document}